\documentclass[12pt]{iopart}
\expandafter\let\csname equation*\endcsname\relax

\expandafter\let\csname endequation*\endcsname\relax
\usepackage{mathrsfs}
\usepackage{mathtools}
\usepackage{physics, tensor}
\usepackage{cite}
\usepackage[margin=1in]{geometry}
\usepackage[linktocpage=true, hidelinks, colorlinks, linkcolor={blue}, urlcolor={cyan}, citecolor={red}]{hyperref}

\begin{document}

\title[GSL for non-minimal theories]{Generalized Second Law for Non-minimally Coupled Matter Theories}

\author{Prateksh Dhivakar$^1$ and Krishna Jalan$^{2,3}$}

\address{$^1$Department of Physics, Indian Institute of Technology Kanpur, Kalyanpur, Kanpur 208016, India }
\address{ $^2$The Institute of Mathematical Sciences, 4th Cross Street, C.I.T. Campus, Taramani, Chennai 600113, India}
\address{$^3$Homi Bhabha National Institute, Training School Complex, Anushakti Nagar, Mumbai 400094, India}
\eads{\mailto{prateksh@iitk.ac.in}, \mailto{krishnajalan@imsc.res.in}}
\vspace{10pt}
\begin{indented}
\item[]September 2023
\end{indented}

\begin{abstract}
We prove the generalized second law (GSL) for higher curvature gravity theories when the matter sector is non-minimally coupled. The validity of our proof is in the regime of linearized fluctuations about equilibrium black holes, which is the same regime as considered in the previous proofs by Wall and Sarkar. These proofs were provided in different gravity theories -- for instance, Lovelock theory and higher curvature gravity -- but the matter sector was always taken to be minimally coupled. In this article, we describe how to generalize the proof of linearized semi-classical GSL when the matter sector comes with non-minimal couplings. The proof proceeds by suitably evaluating the matter path integral in the stress tensor expectation value by treating the higher derivative couplings in an effective field theory setting. We use the recently established result of the linearized second law for such theories.
\end{abstract}



    \newcommand{\act}{\mathcal{S}}
    \newcommand{\lagr}{\mathcal{L}}
    \newcommand{\ric}{\mathcal{R}}
    \newcommand{\pf}{\mathcal{Z}}
    \newcommand{\ea}{\mathcal{W}}
    \newcommand{\uv}{\ell}
    \newcommand{\pid}{\mathscr{D}}
    \newcommand{\pd}{\partial}
    \renewcommand{\d}{\nabla}
    \newcommand{\cmmnt}[1]{}

%
\vspace{2pc}
\maketitle
%
%

\section{Introduction} \label{sec:intro}
    General Relativity (GR) is considered incomplete due to the presence of singularities \cite{Hawking:1970zqf}, which are expected to be resolved by a quantum theory of gravity. At energies significantly below the Planck scale, any UV complete theory of gravity simplifies to Einstein's general relativity with suppressed higher derivative terms \cite{PhysRevLett.55.2656, Zwiebach:1985uq}. Consequently, we can treat gravity as an effective field theory (EFT) and concentrate on the primary quantum corrections at these energy levels \cite{Donoghue:1994dn}. Black holes, as solutions of Einstein's theory, provide a useful context for studying the quantum nature of gravity. In this framework, black holes exhibit thermodynamic properties, such as well-defined notions of energy and temperature \cite{Bardeen:1973gs, PhysRevD.7.2333, Hawking:1975vcx}. The black hole's entropy is linked to its horizon's area \cite{Hawking:1971vc, Hawking:1971tu}, and the area theorem ensures that entropy never decreases, akin to the second law of black hole thermodynamics.

    Due to the presence of higher derivative terms, there's a question about whether black hole thermodynamics holds beyond general relativity. In the presence of higher derivative interactions, the area law typically fails \cite{Visser:1993qa, Visser:1993nu, LopesCardoso:1999cv, Jacobson:1993xs}. Iyer and Wald, \cite{Wald:1993nt, Iyer:1994ys}, introduced a major generalization for diffeomorphism invariant theories of gravity, where black hole entropy is derived from the integral of the Noether charge associated with the horizon's generating Killing isometry. While Wald entropy satisfies the first law of black hole mechanics, it suffers from JKM ambiguities for non-dynamical horizons  \cite{Jacobson:1993vj, Jacobson:1995uq}. As a result, except for exceptional cases like $f(\ric)$ theories, where $\ric$ is the Ricci scalar and $f$ is a polynomial, Wald entropy doesn't generally adhere to the second law of black hole mechanics, which dictates that entropy should increase at each stage of evolution

    To prove the second law of black hole mechanics, one common approach is to consider an equilibrium black hole with a Killing horizon being slightly perturbed by external matter. As the black hole eventually returns to equilibrium, an entropy functional should be constructed to increase throughout this dynamical process. This can be achieved by extending the Wald entropy to dynamic situations while addressing the associated JKM ambiguities to ensure compliance with the second law. This approach has been used in previous studies \cite{Sarkar:2013swa, Bhattacharjee:2015yaa, Wall:2015raa, Bhattacharjee:2015qaa, Bhattacharya:2019qal, Bhattacharyya:2021jhr, Biswas:2022grc}.

    In \cite{Bhattacharya:2019qal, Bhattacharyya:2021jhr, Biswas:2022grc}, building on \cite{Wall:2015raa}, an ultra-local version of the second law of black hole thermodynamics was established for arbitrary diffeomorphism invariant theories of gravity. This was accomplished by introducing a local ``entropy current'' with a non-negative divergence on the black hole's dynamic horizon. The time component of the entropy current was a combination of the Wald entropy and uniquely resolved JKM ambiguities. The spatial components of the current facilitated the redistribution of entropy across different spatial sections of the horizon. In previous works \cite{Sarkar:2013swa, Wall:2015raa, Bhattacharya:2019qal, Bhattacharyya:2021jhr}, it was assumed that the matter sector of the theory was minimally coupled and satisfied the null energy condition (NEC), which contributed to entropy production and the establishment of the second law. The minimal coupling prescription that we work with is the following: given any Lorentz invariant expression involving the matter fields in flat Minkowski space $\eta_{ab}$, the minimally coupled expression is given by taking the flat metric $\eta_{ab}$ to the curved space metric $g_{ab}$ and the partial derivative $\partial_{m}$ to the covariant derivative $D_{m}$.

    It is well known that if one considers a matter sector non-minimally coupled to gravity, it can violate NEC \cite{Wall:2018ydq, PhysRevD.54.6233, Barcelo:2000zf, Chatterjee:2015uya}. The analysis of \cite{Biswas:2022grc} lifts this crucial assumption, establishing a linearized second law for arbitrary matter couplings to gravity. The current work aims to extend the analysis of \cite{Biswas:2022grc} by treating matter fields quantum mechanically, thereby establishing a Generalized Second Law (GSL).

    The GSL states that the total entropy of the matter fields outside the horizon $S_{\text{out}}$ and the entropy of the black hole $S_{H}$ always increases in a dynamical evolution. Thus, the proof of the GSL proceeds by showing that the generalized entropy $S_{\text{gen}} = S_{H} + S_{\text{out}}$ always increases. 
    
    In the semi-classical setting that we are interested in, one expands the metric in powers of the Planck mass, or equivalently, an $\hbar$ expansion \cite{Wall:2009wm,Wall:2011hj,Sarkar:2013swa}:
    \begin{equation}
        g_{ab} = g^{(0)}_{ab} + g^{({\frac{1}{2}})}_{ab} + g^{(1)}_{ab} + \mathcal{O}(\hbar^{\frac{3}{2}})\, .
    \end{equation}
    The zeroth-order term denotes the background metric, and the half-order term denotes the fluctuations due to quantized gravitons. The first-order term arises from the gravitational field due to quantized matter fields. Since we are considering linearized fluctuations, we will ignore all terms higher than $\mathcal{O}(\hbar)$. As detailed in \cite{Wall:2011hj}, we are interested in a regime where the metric fluctuations are solely driven by the quantum fluctuations of matter fields. This is the most interesting case because if we consider classical fluctuations, they far outweigh the quantum fluctuations. Unlike the previous proofs of GSL in different kinds of gravity theories, we consider the matter sector to be non-minimally coupled. We were able to establish a statement of linearized semi-classical GSL for the non-minimally coupled quantum field in an EFT sense (cf. Section \ref{sec:setup}). In this EFT treatment, we assume a clear separation of scale between the minimal and non-minimal sectors, which results in the splitting of the minimal and non-minimal parts of the stress-energy tensor. We find that the non-minimal sector, treated perturbatively over the minimal sector, should only contribute to the horizon entropy, whereas the minimal sector is the dominant contribution to $S_{\text{out}}$ at the linear order. With this simplification, we can make a statement of GSL for the non-minimally coupled matter by suitably building up on the previous works of Sarkar and Wall \cite{Sarkar:2013swa, Wall:2015raa} and using the boost-weight analysis and entropy current structure of \cite{Bhattacharyya:2021jhr, Biswas:2022grc}. We should stress that this analysis does not incorporate the graviton contributions. We would like to mention that the GSL was considered for a particular example of non-minimal coupling in \cite{Ford_2001}, and it has also been considered for specific examples of pure gravity theories in \cite{Wu:2008ir}.

   Recently, in \cite{Ali:2023bmm}, the authors prove a statement of linearized GSL for higher curvature gravity (including graviton contributions), generalizing the works in \cite{Chandrasekaran:2022cip, Chandrasekaran:2022eqq} (CPLW, CPW). Following up on the work of Leutheusser and Liu \cite{Leutheusser:2021qhd, Leutheusser:2021frk}, Witten in \cite{Witten:2021unn} first showed that there is a change in the nature of the von Neumann algebra associated to local observables in the black hole exterior in AdS spacetimes with holographic boundary duals. This change in the nature of local observable algebra allows for associating a notion of density matrix and a (renormalized) von Neumann entropy with the black hole state, which is well-defined up to a state-independent additive constant. Later in \cite{Chandrasekaran:2022cip, Chandrasekaran:2022eqq}, this idea was generalized to black holes in asymptotically flat spacetimes, such that the entropy of any semi-classical state of the transformed algebra agrees with the generalized entropy, up to an additive constant that is independent of the choice of state. In \cite{Ali:2023bmm}, the authors claim a generalization of the results of \cite{Chandrasekaran:2022cip, Chandrasekaran:2022eqq} to an arbitrary diffeomorphism invariant theory of gravity. However, the monotonicity of the generalized entropy, which is essential to the proof of GSL, was a statement in the boundary theory in \cite{Chandrasekaran:2022eqq}, and it also does not generalize to out-of-equilibrium horizon cuts, unlike the case in \cite{Wall:2011hj} where Wall proves that for any horizon cut, the statement of GSL is given by $ \dv{S_{\text{gen}}}{v} \geq 0 \, $. The main novelty of our proof of GSL is that one does not need the input from the boundary theory, and we also have a statement for arbitrary out-of-equilibrium cuts.

    We conclude this section by briefly summarizing the different sections. In Section \ref{sec:CSL}, we briefly summarize the proof of classical second law (CSL), introducing the idea of horizon-adapted coordinates, boost-weight analysis, and the entropy current structure. These form some key ingredients in our proof for the GSL as well. Readers familiar with these ideas can directly jump to Section \ref{sec:GSL} which is the key section of the paper. In Section \ref{sec:GSL}, we first summarize previous attempts at proving GSL in different gravity theories. This sets the stage to explain our setup and how our proof compares to the previous literature. In this section, we also describe our EFT treatment of the non-minimal sector and give a proof of the linearized semi-classical GSL in the context of non-minimally coupled matter sector. Finally, in Section \ref{sec:discussion}, we discuss the relevance of our work, mentioning some issues to be addressed and pointing out some open questions for future work.

\section{Brief review of the classical second law} \label{sec:CSL}
    In this section, we will recap the basic setup of the proof of linearized second law \cite{Wall:2015raa,Bhattacharya:2019qal,Bhattacharyya:2021jhr,Biswas:2022grc}. We begin by explaining our coordinate choice for the black hole spacetime. Then, we detail the notion of stationarity in our spacetime, allowing us to define the fluctuating quantities about equilibrium solutions. These fluctuating quantities can be classified according to how they transform under the symmetries of the stationary spacetime. One can then argue for the linearized second law from the structures of these fluctuating quantities. We will state the main facts we will use, and we refer \cite{Bhattacharyya:2021jhr} for further details.

    \subsection{Horizon adapted coordinates} \label{sec:gauge}
    The choice of coordinates for our black hole spacetime is given by $\{ v,\ r,\ x^A\}$ where $(A = 1, \ldots, d-2)$ such that the horizon is located at $r = 0$. The coordinate $v$ is chosen along the null generators of the horizon $\partial_v$. On each constant $v$ slice, we shoot off a set of spatial tangents $\partial_A$ in the $d-2$ spatial directions. These spatial tangents $\partial_A$ will span the coordinates $x^A$. Thus, the horizon will be spanned by $\{v,x^A\}$. The spacetime away from the horizon will be spanned by a set of null geodesics $\partial_r$ emanating from the horizon, making appropriate angles with $\partial_A$ and $\partial_v$. This results in the gauge choice of \cite{Bhattacharyya:2021jhr, Biswas:2022grc}, namely
    \begin{equation} \label{eq:gauge}
        \dd s^2 = g_{ab} \dd x^a \dd x^b, \qq{where} g_{rr} = g_{vA} \big|_{r=0} = g_{vv} \big|_{r=0} = g_{rA}  = \pd_r g_{vv} \big|_{r=0} = 0 \, , g_{rv} = 1 \, .
    \end{equation}
    With this, the near-horizon metric for a generic dynamical black hole can always be cast in the form given by \cite{Bhattacharyya:2016xfs}
    \begin{equation} \label{eq:metric}
        \dd s^2 = 2 \dd v \, \dd r - r^2 X(r, v, x^C) \dd v^2 + 2 r \omega_A(r, v, x^C) \dd v \, \dd x^A + h_{AB}(r, v, x^C) \dd x^A \, \dd x^B\, .
    \end{equation}
    We assume a form of the Zeroth law that enables us to express the metric in this form. A Zeroth law for arbitrary pure gravity diffeomorphism invariant theories was established in \cite{Bhattacharyya:2022nqa} following upon initial work in \cite{Ghosh:2020dkk}. 

    Here $X,\omega_A,h_{AB}$ are general functions of coordinates. This gauge choice does not completely fix the choice of coordinates on constant $v$ and $r$ slices as there are residual symmetries in the gauge of \eqref{eq:metric} of the form
    \begin{equation}\label{eq:ressymmetries}
        \begin{split}
            v \to \Tilde{v} &= v f_1(x^i) +f_2(x^i) \, ,\ \text{with appropriate redefinition of} ~r \, , \\
            x^i \to \Tilde{x}^i &= g^i(x^j) \, .
        \end{split}
    \end{equation}

    We work within a regime where we are perturbatively close to the equilibrium solution. For this, we must construct an equilibrium solution in our gauge \eqref{eq:metric}. We will use the residual symmetries of \eqref{eq:ressymmetries} to facilitate this. Let us consider a simple scaling symmetry of the form
    \begin{equation} \label{eq:boost_transformation}
        (r,v) \mapsto \left( \lambda r, \frac{v}{\lambda} \right) \, .
    \end{equation}
    This is called ``boost transformation'' and is a particular case of \eqref{eq:ressymmetries} when $f_1$ is a constant and $f_2=0$. Since equilibrium black holes must have a Killing vector that becomes null on the horizon, we can use \eqref{eq:boost_transformation} to fix our stationary background metric. As \eqref{eq:boost_transformation} is generated by
    \begin{equation} \label{eq:boost_generator}
        \xi = v \pd_v - r \pd_r \, ,
    \end{equation}
    a stationary black hole with metric $g_{ab}^{\text{(eq)}}$ is given by
    \begin{equation} \label{eq:horizon_adapated_gauge}
        \dd s^2 = 2 \dd v \dd r - r^2 X(rv, x^C) \dd v^2 + 2 r \omega_A(rv, x^C) \dd v \dd x^A + h_{AB}(rv, x^C) \dd x^A \dd x^B,
    \end{equation}
    where the coefficients $X, \omega_A, h_{AB}$ are now functions of the products of the coordinates $r, v$. See \ref{ap:examplegauge} for a demonstration of how the standard Schwarzschild and Kerr solutions can be brought to this gauge. The metric is invariant under the scaling \eqref{eq:boost_transformation}. This is said to be the boost symmetry of the horizon, and \eqref{eq:boost_generator} is a Killing vector of the background spacetime. The horizon $r=0$ is generated by \eqref{eq:boost_generator} and is a Killing horizon of the spacetime. $\xi^{\mu}$ of \eqref{eq:boost_generator} becomes null on the horizon $r=0$. For our construction, it is enough to consider a Killing horizon. We will assume that the event horizon of the equilibrium black hole is a Killing horizon. We want to describe the dynamics only in a perturbative sense, up to the linearized order, in which case we decompose the metric $g_{ab}$ in \eqref{eq:metric} using \eqref{eq:horizon_adapated_gauge} as
    \begin{equation} \label{eq:linear_order_metric}
        g_{ab} = g^{\text{(eq)}}_{ab}(rv, x^C) + \delta g_{ab}(r, v, x^C),
    \end{equation}
    where $g^{\text{(eq)}}_{ab}$ is the equilibrium metric as in \eqref{eq:horizon_adapated_gauge}, and $\delta g_{ab}$ is the fluctuation in the metric at the linearized order in $\hbar$ or $\epsilon$.

    Once we have the notion of the background plus fluctuation, we can quantify non-equilibrium structures based on their ``boost weight''. The boost weight is defined according to how covariant tensors/structures transform under the boost transformation \eqref{eq:boost_transformation}:
    \begin{equation}
        \mathcal{T} \to \bar{\mathcal{T}} = \lambda^w \mathcal{T} \, , ~ \text{under} ~ \left\{ r \to \bar{r} = \lambda r, \, v \to \bar{v} = \frac{v}{\lambda} \right\} ~~ \implies ~~ \text{boost weight of $\mathcal{T}$ is $w$} \, .
    \end{equation}

    We have a non-equilibrium quantity if $w > 0$ for a covariant tensor. This can be understood as follows. Suppose if $\mathcal{T}$ was of the form
    \begin{equation}\label{eq:flucT}
        \mathcal{T} \sim (\partial_r)^{k_r} (\partial_v)^{k_v} \mathcal{A} \, ,
    \end{equation}
    where $\mathcal{A}$ is constructed out of $X,\omega_A,h_{AB}$ and $\nabla_A$ (the covariant derivative compatible with induced metric $h_{AB}$) only. When we evaluate $\mathcal{A}$ on \eqref{eq:linear_order_metric}, it breaks up into an equilibrium contribution and a fluctuating contribution. The equilibrium contribution $\mathcal{A}(r,v,x^A) \big|_{\text{eqbrm}} \sim \mathcal{A}(rv,x^A)$. If we take $k_v > k_r$, i.e., $w>0$, then the equilibrium contribution of \eqref{eq:flucT} has $k_v-k_r$ factors of $r$. This evaluates to zero on the horizon $r=0$. Thus, only the fluctuating part of \eqref{eq:flucT} remains:
    \begin{equation}
        \mathcal{T} \big|_{r=0} \sim (\partial_r)^{k_r} (\partial_v)^{k_v} \mathcal{A} \sim \epsilon (\partial_r)^{k_r} (\partial_v)^{k_v} \delta \mathcal{A} \sim \mathcal{O}(\epsilon) \, .
    \end{equation}
     Now suppose if we consider a product of two such terms, then they become $\mathcal{O}(\epsilon^2)$ and we neglect them up to the linearized order we are working with:
     \begin{equation}\label{eq:quadboostwt}
         (\partial^{m_1}_r \partial^{m_1+k_1}_v) \mathcal{A}_1 \, (\partial^{m_2}_r \partial^{m_2+k_2}_v) \mathcal{A}_2 \sim \mathcal{O}(\epsilon^2) \, .
     \end{equation}
     Thus, when we have positive boost weight quantities, they are non-trivial only for non-equilibrium configurations. Likewise, we neglect the terms, which are in the form of a product of two positive boost weight quantities at the linearized approximation we are working with.
     
    \subsection{Review of the second law within the setup of the entropy current}
    The entropy of a dynamical black hole in arbitrary diffeomorphism invariant theories is given as \cite{Sarkar:2013swa,Wall:2015raa,Bhattacharya:2019qal,Bhattacharyya:2021jhr}
    \begin{equation} \label{eq:total_entropy}
        S_{\text{tot}, v} = \frac{1}{4\hbar G} \int_{H_v} \dd^{d-2} x\, \sqrt{h}\, (1+ s^{\text{HD}}_{\text{wald}} + s_{\text{cor}})\, ,
    \end{equation}
    where, $h_{AB}$ is the induced metric on the space-like $(d-2)$ dimensional cross-section of the horizon $H_v$, (the subscript $v$ indicates we are looking at some constant $v$ slice) and $h$ denotes its determinant. The factor of 1 contribution arises from the two-derivative Einstein-Hilbert term in the gravity Lagrangian and it is the standard Area term, $s^{\text{HD}}_{\text{wald}}$ is the entropy density contribution arising only from the higher-derivative terms in the gravity action, and $s_{\text{cor}}$ is the possible correction to the entropy density due to non-equilibrium effects, which includes JKM ambiguities. It is not contained in the Wald entropy density $s_{\text{wald}} = 1 + s^{\text{HD}}_{\text{wald}}$. Clearly, in the equilibrium limit, this contribution should vanish, i.e., $s_{\text{cor}} \big|_{\text{eqbrm}} \to 0$.

    Define $\rho = s_{\text{wald}} + s_{\text{cor}}$. As we perturb the horizon, the entropy changes and the rate of change is given by
    \begin{equation} \label{eq:entropy_change_rate}
        \dv{S_{\text{tot}, v}}{v} = \frac{1}{4\hbar G} \int_{H_v} \dd^{d-2} x\, \sqrt{h}\, \left[\dv{\rho}{v} + \theta \rho \right]\, ,
    \end{equation}
    where $\theta$ is the expansion parameter of the null generators and is given by
    \begin{equation} \label{eq:expansion_parameter}
        \theta = \frac{1}{\sqrt{h}} \partial_v \sqrt{h} \, .
    \end{equation}
    The total change can then be obtained as
    \begin{equation} \label{eq:entropy_change}
        \Delta S_{\text{tot}} = \frac{1}{4\hbar G} \int_H \dd v\, \dd^{d-2} x\, \sqrt{h}\, \left( \dv{\rho}{v} + \theta \rho \right)\, .
    \end{equation}
      We now define the ``generalized expansion parameter'' $\Theta$ as the rate at which the entropy density of an infinitesimal region of the horizon changes, i.e.,
    \begin{equation} \label{eq:expansion_parameter_generalized}
        \Theta \coloneqq  \dv{\rho}{v} + \theta \rho\, ,
    \end{equation}
    which for the Einstein-Hilbert term reduces to the usual expansion parameter $\theta$ that goes into the Raychaudhuri equation \cite{Raychaudhuri:1953yv}. It will be assumed that the dynamical horizon will eventually reach a stationary configuration such that $\Theta(v \to \infty) = 0$. Thus, the classical second law follows if we can show $\Delta S_{\text{tot}} > 0$. A way to show this is to prove $\Theta > 0$, which is a stronger statement than $\Delta S_{\text{tot}} > 0$. This follows if we prove $\partial_v \Theta < 0$ and then use $\Theta(v \to \infty) = 0$ to infer $\Theta > 0$ for every finite $v$. 

    Now $\partial_v \Theta$ is given by
    \begin{equation}\label{eq:delvtheta}
        \begin{split}
            \partial_v \Theta &= \partial^2_v (s^{\text{HD}}_{\text{wald}} + s_{\text{cor}}) + \partial_v \theta (1+ s^{\text{HD}}_{\text{wald}} + s_{\text{cor}}) + \theta \partial_v \rho \\ &= \partial_v \theta + \partial_v \left( \dfrac{1}{\sqrt{h}} \partial_v \left( \sqrt{h} (s^{\text{HD}}_{\text{wald}} + s_{\text{cor}}) \right) \right) + \mathcal{O}(\epsilon^2) \\
            &= - \ric_{vv} + \partial_v \left( \dfrac{1}{\sqrt{h}} \partial_v \left( \sqrt{h} ( s^{\text{HD}}_{\text{wald}} + s_{\text{cor}} ) \right)  \right) + \mathcal{O}(\epsilon^2) \, .
        \end{split}
    \end{equation}
    In the second equality, we have used the boost weight prescription of \eqref{eq:quadboostwt} to get rid of quadratic terms like $\theta \partial_v \rho$ and other terms that arise as the result of expressing them in the above form. In the final equality, we have used the linearized Raychaudhuri equation to write $\partial_v \theta = - \ric_{vv} + \mathcal{O}(\epsilon^2)$. Since the Lagrangian is given by the Einstein-Hilbert (Ricci scalar) term in addition to higher derivative terms and a matter sector (non-minimally coupled matter plus a minimally coupled sector that satisfies the NEC), the equations of motion projected onto $r=0$ (the horizon) in our gauge \eqref{eq:metric} are given by
    \begin{equation}\label{eq:eomsubs}
        \ric_{vv} + E^{\text{HD}}_{vv} + T^{\text{(non-min)}}_{vv} = T^{\text{(min)}}_{vv} \, .
    \end{equation}
    Here $E_{vv} = \ric_{vv} + E^{\text{HD}}_{vv}$ denotes the contribution of pure gravity terms, and $T_{vv}$ denotes the contribution of matter terms coupled to gravity tensors. Substituting \eqref{eq:eomsubs} in \eqref{eq:delvtheta}, we get
    \begin{equation}\label{eq:delvthetafinal}
        \partial_v \Theta = E^{\text{HD}}_{vv} + T^{\text{(non-min)}}_{vv} + \partial_v \left( \dfrac{1}{\sqrt{h}} \partial_v \left( \sqrt{h} ( s^{\text{HD}}_{\text{wald}} + s_{\text{cor}} ) \right)  \right) - T^{\text{(min)}}_{vv} + \mathcal{O}(\epsilon^2) \, .
    \end{equation}
    From \eqref{eq:delvthetafinal}, it is clear that if one can show that the equations of motion projected onto the horizon $r=0$ in our gauge \eqref{eq:metric} satisfy
    \begin{equation}\label{eq:Evvstruc}
        E^{\text{HD}}_{vv} + T^{\text{(non-min)}}_{vv} = - \partial_v \left( \dfrac{1}{\sqrt{h}} \partial_v \left( \sqrt{h} ( s^{\text{HD}}_{\text{wald}} + s_{\text{cor}} ) \right)  \right) + \mathcal{O}(\epsilon^2) \, ,
    \end{equation}
    then we have
    \begin{equation}
        \partial_v \Theta = - T^{\text{(min)}}_{vv} + \mathcal{O}(\epsilon^2) < 0 \, ,
    \end{equation}
    because of null energy condition $T^{\text{(min)}}_{vv} > 0$. It is important to note that this inequality sign in the context of linearized approximation necessarily implies that $T^{\text{(min)}}_{vv}$ cannot be $\mathcal{O}(\epsilon)$.

    This is because one can switch the sign of $\epsilon$ to violate the inequality. Thus, we have $T^{(\text{min})}_{vv} \sim \mathcal{O}(\epsilon^2)$ and we are essentially proving $\partial_v \Theta \sim \mathcal{O}(\epsilon^2)$. An intuitive way to understand this is that there are no terms in the linear order that can make $\partial_v \Theta > 0$. Such terms might occur in the quadratic order, and one must carefully analyze them like \cite{Hollands:2022fkn}. We will not have anything to say about the non-linear order in our paper. 

    In the above proof, the structure of equations of motions projected onto the null horizon in \eqref{eq:Evvstruc} played a crucial role. Consider a Lagrangian which is a diffeomorphism invariant theory of gravity of the following form
    \begin{equation}\label{eq:lagrangian}
        \lagr = \lagr(g_{ab}, \ric_{abcd}, D_{e} \ric_{abcd}, \dots, \Phi, D_a \Phi, D_{(a}D_{b)} \Phi, \dots, F_{ab} , D_{c} F_{ab},\dots) \, ,
    \end{equation}
    where the Lagrangian can have a complicated dependence of the metric $g_{ab}$, scalar $\Phi$, and $U(1)$ gauge fields $A_{a}$, with field strength tensor $F_{ab}$. If one can show that the equations of motion projected onto the null horizon $r=0$ in our gauge \eqref{eq:metric} take the form
    \begin{equation}\label{eq:Evvfinal}
         E^{\text{HD}}_{vv} + T^{\text{(non-min)}}_{vv} =  \partial_v \left( \dfrac{1}{\sqrt{h}} \partial_v \left( \sqrt{h} \mathcal{J}^v \right) + \nabla_A \mathcal{J}^A  \right)  + \mathcal{O}(\epsilon^2) \, ,
    \end{equation}
    then \eqref{eq:Evvstruc} holds, and thus, a linearized second law holds for black holes in the theory we are considering. \eqref{eq:Evvfinal} has been established for the Lagrangian of the form \eqref{eq:lagrangian} in \cite{Bhattacharyya:2021jhr,Biswas:2022grc}. Using constraints from the boost symmetry \eqref{eq:boost_transformation} of the horizon and diffeomorphism invariance of the Lagrangian, \cite{Bhattacharyya:2021jhr} showed that $E_{vv}$ of \eqref{eq:eomsubs} has the structure of \eqref{eq:Evvfinal}. The input of diffeomorphism invariance can be relaxed as this structure was also established for Chern-Simons theories of gravity, which are diffeomorphism invariant up to total derivatives only \cite{Deo:2023vvb}. The analysis follows by carefully working out how different covariant tensors, built out of the Lagrangian, transform under the boost transformation \eqref{eq:boost_transformation}. For $T_{vv}^{\text{(non-min)}}$, we can invoke the analysis of \cite{Biswas:2022grc} to show that it also has the structure of \eqref{eq:Evvfinal}. This proves the linearized classical second law. Here, $\mathcal{J}^v$ and $\mathcal{J}^A$ are considered as components of a $d-1$ ``vector'' on the horizon known as entropy current. $\mathcal{J}^v$ is given by
    \begin{equation}
        \mathcal{J}^v = 1 + s^{\text{HD}}_{\text{wald}} + s_{\text{cor}} \, ,
    \end{equation}
    which is the equilibrium Wald entropy density plus the associated JKM ambiguities, which are now fixed as we are just rewriting the equations of motion. We put vector in quotes because as the horizon is null, this is not a covariant vector of the full space-time. If we integrate $\mathcal{J}^v$ over the cross-section of the horizon, we get the total entropy \eqref{eq:total_entropy}. $\mathcal{J}^A$ is a non-equilibrium quantity, and it can be understood as quantifying the redistribution of entropy across the spatial cross-section of the constant $v$ slices. This interpretation holds because the increase of the expansion parameter $\Theta >0$ implies
    \begin{equation}
        \dfrac{1}{\sqrt{h}} \partial_v \left( \sqrt{h} \mathcal{J}^v \right) + \nabla_A \mathcal{J}^A > 0 \, .
    \end{equation}
    We have the result that the divergence of an entropy current with components $\mathcal{J}^v$ and $\mathcal{J}^A$ is positive. This is a stronger ultra-local version of the linearized second law when compared to the integrated version of the second law considered in \cite{Wall:2015raa}, i.e., in \eqref{eq:entropy_change}. This is because the divergence of entropy current not only quantifies an increase in entropy along the ``time'' direction but also quantifies a redistribution along the spatial directions in such a way that the total entropy always increases (because there is ultra-local entropy production). Clearly, the presence of $\mathcal{J}^A$ would not contribute to \eqref{eq:entropy_change_rate} if we assume compact horizons, whence the total derivative term drops out of the spatial integral. Thus, \eqref{eq:Evvfinal} and \eqref{eq:Evvstruc} are equivalent. At this point, one might raise an objection that the construction is heavily reliant on the horizon adapted coordinates. However, one can show that the entropy current structure is covariant under affine reparametrizations of the null generators \cite{Bhattacharyya:2022njk}.  This completes our review of the linearized classical second law.

\section{The Generalized Second Law} \label{sec:GSL}
    In this section, we will give a proof of the GSL when the matter is non-minimally coupled to gravity. We work in an EFT sense, which will be elaborated in Section \ref{sec:setup}. This also requires defining the expectation value of the stress-energy tensor, which involves the first-order quantum correction to the classical theory and will be treated similarly as in \cite{Birrell:1982ix}. We then give a proof of the linearized semi-classical GSL in Section \ref{sec:proof}.
    
    \subsection{Summary of GSL in other gravity theories} \label{sec:review}
    We briefly summarize \cite{Sarkar:2013swa, Wall:2015raa}, which formed the primary motivation for our work. Sarkar and Wall (SW) \cite{Sarkar:2013swa} constructed a proof for the (integrated form of) linearized classical second law and the linearized generalized second law for the semi-classical, minimally-coupled matter sector in the context of Lovelock theories. This was later extended to an arbitrary diffeomorphism invariant theory of gravity with minimally coupled matter by Wall \cite{Wall:2015raa}. The setup considered has been detailed in Section \ref{sec:CSL}: a stationary black hole is perturbed slightly and it settles down to some equilibrium solution. This allows us to use the linearized approximation in the dynamics. The central idea was to argue that \eqref{eq:total_entropy} with the JKM ambiguities fixed, resulted in a linearized version of the second law. 
    
    The semi-classical equation of motion for the metric is given by
    \begin{equation} \label{eq:EoM_metric}
        E_{ab} = 8\pi G_N \ev{T^{\text{(full)}}_{ab}} \, ,
    \end{equation}
    where $E_{ab}$ will be considered to be a c-number whereas the stress-energy tensor $T^{\text{(full)}}_{ab}$ is treated as a quantum operator. The precise meaning of $\ev{T^{\text{(full)}}_{ab}}$ will be described in Section \ref{sec:stress_tensor_expectation}, \cite{Birrell:1982ix, Wall:2011hj}. The state with respect to which the expectation value is calculated is chosen to be the generic semi-classical vacuum state corresponding to the background equilibrium solution $g_{ab}^{(0)}$. The semi-classical equation in \eqref{eq:EoM_metric} is justified because we will be working in the linearized weak-field approximation limit, where the semi-classical equation in \eqref{eq:EoM_metric} is obtained from the full quantum operator equation by taking expectation values of the $\mathcal{O}(\hbar)$ part of the metric, \cite{Wall:2009wm}. In \cite{Sarkar:2013swa}, $T^{\text{(full)}}_{ab}$ received contribution just from the minimally coupled matter sector, but for \cmmnt{us}a generic non-minimal matter theory it would include the non-minimal terms as well. Then, if we substitute \eqref{eq:EoM_metric} instead of \eqref{eq:eomsubs} in \eqref{eq:delvtheta}, we have for first-order changes
    \begin{equation} \label{eq:change_expansion_parameter_generalized}
        \dv{\Theta}{v} \approx - 8 \pi G_N \ev{T^{\text{(full)}}_{vv}}  \, . 
    \end{equation}
    Now, the claim is that in any theory of gravity, if \eqref{eq:change_expansion_parameter_generalized} is obeyed, then this will imply the linearized semi-classical GSL \cite{Sarkar:2013swa, Wall:2015raa}. This can be seen as follows. We first integrate \eqref{eq:change_expansion_parameter_generalized} once in the transverse directions $x^A$ and twice along the null generator $\partial_v$ to get
    \begin{equation}\label{eq:final_GSL_stat}
        \dfrac{\hbar}{2 \pi} \left( S_{\text{tot}}(\infty) - S_{\text{tot}}(v') \right) = \int \dd^{d-2} x \sqrt{h} \int_{v > v'} \dd v \, (v - v')  \ev{T^{\text{(full)}}_{vv}} \, .
    \end{equation}
    \cite{Sarkar:2013swa, Wall:2015raa} showed that this would imply a semi-classical GSL for matter minimally coupled to gravity.
    
    In both works above, certain assumptions were made on the algebra of observables on the horizon. These assumptions were first detailed in Wall's earlier work on proving GSL for Einstein's theory with fairly generic fields minimally coupled to gravity and for arbitrary horizon slices, \cite{Wall:2011hj}. All the above proofs required some (as in \cite{Sarkar:2013swa}) or all (as in \cite{Wall:2011hj}) of the four axioms that the algebra of observables restricted to the horizon must satisfy, namely, Determinism (the horizon algebra completely specifies all the information falling across the horizon such that together with the information available at the future null infinity, all the information outside the event horizon is completely determined), Ultralocality (the degrees of freedom on distinct horizon generators should be understood as independent systems), Local Lorentz Invariance (there exists an infinite dimensional group of symmetries, associated with the horizon algebra, which correspond to the affine transformations of each horizon generator), and Stability (the generator of null translations should be non-negative).
    
    In \cite{Ali:2023bmm}, the authors arrive at \eqref{eq:final_GSL_stat} by crucially using the structure of $E_{vv}$ in \eqref{eq:Evvstruc} proved for arbitrary diffeomorphism invariant theories of gravity in \cite{Bhattacharyya:2021jhr, Biswas:2022grc} following up on the works of \cite{Sarkar:2013swa, Wall:2015raa, Chatterjee:2011wj, Mishra:2017sqs}. For instance, (II.38) of \cite{Ali:2023bmm} follows crucially from equation (5.7) of \cite{Bhattacharyya:2021jhr}. They then proceed to generalize the constructions of \cite{Chandrasekaran:2022cip,Chandrasekaran:2022eqq} for arbitrary diffeomorphism invariant theories of gravity by suitably generalizing the notions of canonical energy in the covariant phase space formalism of \cite{Wald:1993nt,Iyer:1994ys}. However, as mentioned in the Introduction (cf. Section \ref{sec:intro}), the proof of GSL 
    does not generalize to out-of-equilibrium cuts as the notion of the crossed product type II von Neumann algebra, and hence an associated entropy, does not exist corresponding to such arbitrary horizon cuts.

    \subsection{The setup and a proof} \label{sec:setup}
    In this section, we will explain in detail the setup in which we prove the GSL. We start with the following action
    \begin{equation} \label{eq:action}
        \begin{split}
            \act &= \act_{\text{g}}[g_{ab}, \ric_{abcd}, D_m \ric_{abcd}, D_{(m} D_{n)} \ric_{abcd}, \ldots] + \act_{\text{min}}[g_{ab}, \Phi, D_m \Phi] \\ &+ \act_{\text{nm}}[g_{ab}, \ric_{abcd}, D_m \ric_{abcd}, D_{(m} D_{n)} \ric_{abcd}, \ldots, \Phi, D_m \Phi, D_{(m}D_{n)} \Phi,\ldots ] \, ,
        \end{split}
    \end{equation}
    where $\act_{\text{g}}$ is just the gravitational part of the action which is an arbitrary function of the $\ric_{abcd}$ and covariant derivatives acting on it, $\act_{\text{min}}$ represents the minimally coupled matter sector (containing terms only up to two derivatives in the field), and $S_{\text{nm}}$ represents the non-minimal interaction of the gravity and matter fields, and we assume that it has to have terms with at least four derivatives in them.

    We want to show that GSL holds to the linear order in $\hbar$ corrections within the EFT approximation of \cite{Hollands:2022fkn}. We will consider the Lagrangian to be a formal sum of different terms with increasing number of derivatives in the fields. Each term comes with a suitable factor of some UV length scale $\uv$, more precisely, a term with $n + 2$ derivatives will be multiplied by $\uv^n$. The terms with two or fewer derivatives are understood as the usual Einstein-Hilbert term with a minimally coupled matter sector, besides some cosmological constant. The validity of EFT depends on the fact that the terms with an increasing number of derivatives become increasingly less significant. We can achieve this by restricting to spacetimes varying over some characteristic length/time scale $L$ with $\frac{\uv}{L} \ll 1$. The scale $L$ could be considered a lower bound on the size of the final equilibrium black hole state and any perturbation scale away from the equilibrium, \cite{Hollands:2022fkn}.

    We will further consider the semi-classical regime where we can control the physical effects as a perturbative expansion in $\hbar$ and $\frac{\uv}{L}$. We will aim to argue for the GSL to linear order in $\hbar$ from \eqref{eq:change_expansion_parameter_generalized}. The non-trivial part of the proof would be to develop a proposal that can evaluate the expectation value of the matter stress tensor when the matter is non-minimally coupled. One cannot straight-away use the four axioms of \cite{Wall:2011hj} to deal with the stress tensor expectation value because non-minimal matter contributes to the horizon entropy. Our proposal in Section \ref{sec:stress_tensor_expectation} will deal with this issue in such a way that we can argue for the linearized GSL from \eqref{eq:final_GSL_stat}. The higher order corrections to this equation in $\hbar$ come from the renormalization theory, but these are safely neglected in our analysis, which treats backreaction only at the leading order in $\hbar$. This is along the same lines as previous analyses of Wall in \cite{Wall:2009wm, Wall:2015raa}, and Sarkar and Wall in \cite{Sarkar:2013swa}.
    
    Further, we stress that $\hbar \sim \epsilon$, which follows from the fact that we treat the metric fluctuations as semi-classical, i.e., the size of the quantum backreaction is comparable to the dynamical fluctuations in the classical background. This is, in fact, the most interesting regime where the validity of GSL should be carefully checked. If we consider classical fluctuations, they far outweigh any quantum fluctuation. Thus, any violation of GSL is apparent only when the size of the classical fluctuations is comparable to the quantum fluctuations \cite{Wall:2011hj}. Thus, we consider the variation of the metric fluctuations along the horizon in the semi-classical case to be comparable to that of the classical fluctuations.

    \subsubsection{Stress tensor expectation value: a proposal} \label{sec:stress_tensor_expectation}
    In this subsection, we will describe a prescription for computing the expectation value of the stress tensor for the complete non-minimal theory by treating the non-minimal part as a small perturbation over the minimal theory, where we know how to compute the expectation values of operators, in particular the stress tensor, up to the one-loop level.\footnote{The theory truncated at one-loop level has all terms of the complete quantum theory to $\mathcal{O}(\hbar)$, and it is in this sense the first order quantum correction to the classical theory, \cite{Birrell:1982ix}. For a discussion on the issues related to semi-classical Einstein equation see \cite{Flanagan:1996gw}.}
    
    The basic object we need to analyze is the generating functional of the full theory given by
    \begin{equation} \label{eq:partition_function}
        \pf[J] = \int \pid \Phi\, \exp[\dfrac{\iota}{\hbar} (\act_{\text{g}} + \act_{\text{min}} + \act_{\text{nm}}) + \iota \int \dd^D x\, J(x) \Phi(x)] \, ,
    \end{equation}
    taken over the space of fields $\Phi$ with suitable measure. Since the $\act_{\text{g}}$ term does not depend on $\Phi$, it will contribute as an overall multiplicative factor to the generating functional, which we denote by $\mathcal{N}_g$. Further, we can analyze the non-minimal part of the action as a perturbation over the minimal theory because $\act_{\text{nm}}$ is at least $\mathcal{O}((\frac{\ell}{L})^2)$ suppressed compared to $\act_{\text{min}}$. Thus we have
    \begin{equation} \label{eq:partition_function_simplified}
        \pf[J] = \mathcal{N}_g \int \pid \Phi\, \exp[\dfrac{\iota}{\hbar}\, \act_{\text{min}} + \iota \int \dd^D x\, J(x) \Phi(x)] \exp[\dfrac{\iota\, \uv^2}{\hbar\,L^2}\act_{\text{non-min}}] \, ,
    \end{equation}
    where $$\act_{\text{nm}} = \frac{\uv^2}{L^2} \act_{\text{non-min}} \, .$$
    We now define the ``minimal generating function'' as
    \begin{equation} \label{eq:minimal_partition_function}
        \pf_{\text{min}}[J] = \int \pid \Phi\, \exp[\dfrac{\iota}{\hbar}\, \act_{\text{min}} + \iota \int \dd^D x\, J(x) \Phi(x)]\, ,
    \end{equation}
    and for any operator in the matter sector, $\hat{A}$, defined in the full non-minimal theory, we define
    \begin{equation} \label{eq:expectation_value}
        \ev{\hat{A}(\Phi, D_m \Phi, \ldots)} \coloneqq \frac{\int \pid \Phi\, e^{\frac{\iota}{\hbar}\, \act_{\text{min}}} A(\Phi, D_m \Phi, \ldots)}{\int \pid \Phi\, e^{\frac{\iota}{\hbar}\, \act_{\text{min}}}} \, .
    \end{equation}
    This implies \eqref{eq:partition_function} takes the form
    \begin{equation}\label{eq:nonminimal_partition}
     \pf[J] = \pf_{\text{min}}[0] \, \ev{\exp[ \dfrac{\iota \, \uv^2}{\hbar \, L^2} \, \act_{\text{non-min}}]} \, .
    \end{equation}
    At this point, it is helpful to comment on how we treat the gravity tensors in the path integral. At the level of the matter path integral, they are quantities independent of the matter fields. Since we have not imposed any gravity equations of motion, we can directly evaluate the path integral over the matter fields by treating gravity tensors as some ``background'' quantities. After the path integral, the matter fields couple to gravity through the semi-classical equation \eqref{eq:EoM_metric}. Only at this point, the geometric tensors would couple to quantum matter fields non-trivially.

    The basic idea of the calculation is to mimic the interaction theory analysis for perturbative QFT. Ultimately, it boils down to calculating things with respect to the minimal theory in the sense that the weight factor in the path integral is defined only via the minimal part. Our EFT expansion is based on the parameter
    \begin{equation}\label{eq:eftparameter}
        \lambda = \dfrac{\ell^2}{\hbar L^2} \ll 1 \, .
    \end{equation}
    This parameter clarifies why we can safely neglect renormalization effects on the couplings. They tend to be suppressed because, at one loop, we have $\lambda \hbar \ll \lambda \ll 1$. We then define a quantity which we call the \textit{effective action} for the quantum matter fields as
    
    \begin{equation} \label{eq:effective_action}
        \ea_{\text{min}} = -\iota \hbar \, \log \pf_{\text{min}}[0] \equiv -\iota \hbar \, \log \pf_{\text{min}} \, ,
    \end{equation}
    and
    \begin{equation}
        \ea = -\iota \hbar \log \pf[0] = -\iota \hbar \log \left[\pf_{\text{min}} \ev{\exp[ \iota \lambda \, \act_{\text{non-min}}]} \right] \, ,
    \end{equation}
    \vspace{10pt}
    such that
    
    \begin{equation} \label{eq:stress_tensor}
        \dfrac{2}{\sqrt{-g}}\fdv{\ea_{\text{min}}}{g_{ab}} \coloneqq \ev{T_{\text{min}}^{ab}} = -\frac{2\iota\, \hbar}{\sqrt{-g}\pf_{\text{min}}} \fdv{\pf_{\text{min}}}{g_{ab}} \qq{and} \dfrac{2}{\sqrt{-g}}\fdv{\ea}{g_{ab}} \coloneqq \ev{T^{ab}} \, .
    \end{equation} 
    \vspace{10pt}
    
    So the expectation value of the stress tensor is defined to be
    \begin{equation}\label{eq:Tab_new}
        \ev{T^{ab}} = \ev{T^{ab}_{\text{min}}} - \dfrac{2\iota}{\sqrt{-g}} \dfrac{\hbar}{\ev{\exp[\iota \lambda \, \act_{\text{non-min}}] }} \fdv{g_{ab}} \ev{\exp[\iota \lambda \,\act_{\text{non-min}}] } \, .
    \end{equation}
    
    We emphasize that \cite{Birrell:1982ix} gives us a prescription for defining the expectation value of the minimal sector only. The above expectation value for a generic non-minimal coupling is our prescription based on EFT, where we crucially required that $\act_{\text{non-min}}$ be suppressed compared to $\act_{\text{min}}$. To evaluate the final step, we first note that the expectation value is defined with respect to the minimal theory in \eqref{eq:expectation_value}. We can simplify the path integral by doing a saddle point approximation. This amounts to setting the field $\Phi$ in $\act_{\text{non-min}}$ as the field $\Phi_{\text{min}}^{\text{sol}}$ which satisfies the classical minimal equation of motion, i.e.,
    \begin{equation} \label{eq:minimal_solution}
        D_a D^a \Phi_{\text{min}}^{\text{sol}} + m^2 \Phi_{\text{min}}^{\text{sol}} = 0\, .
    \end{equation}
    This saddle-point approximation is valid because the minimal part of the action has only up to second-order derivatives in the field. Thus, we have
    \begin{equation}\label{eq:expectationnonmin_final}
        \ev{\exp[\iota \lambda \, \act_{\text{non-min}}] } = \exp[\iota \lambda \, \act_{\text{non-min}}[\Phi= \Phi^{\text{sol}}_{\text{min}} ]] \, .
    \end{equation}
    We stress the point that we were able to arrive at the above equation because $\act_{\text{non-min}}$ is suppressed compared to $\act_{\text{min}}$ in an EFT sense \eqref{eq:eftparameter}. Using \eqref{eq:expectationnonmin_final} in \eqref{eq:Tab_new}, we get \vspace{10pt}
    \begin{equation}
        \ev{T^{ab}} = \ev{T^{ab}_{\text{min}}} + \dfrac{2 \lambda \, \hbar}{\sqrt{-g}} \fdv{g_{ab}}  \act_{\text{non-min}}[\Phi = \Phi^{\text{sol}}_{\text{min}}] = \ev{T^{ab}_{\text{min}}} + \dfrac{\uv^2}{L^2} \,  T^{ab}_{\text{non-min}}[\Phi = \Phi^{\text{sol}}_{\text{min}}] \, .
    \end{equation}

    \vspace{10pt}
    For our proof, we would be interested only in the $vv$-component of the  stress tensor, using the fact that the terms are either proportional to $g_{vv}$,
    or have positive boost weight, and with our gauge choice \eqref{eq:horizon_adapated_gauge} it is easy to see that
    \begin{equation} \label{eq:minvv_component_at_horizon}
        \ev{T_{vv}^{\text{(min)}}}\big|_{r=0} \sim \mathcal{O}(\hbar) \, .
    \end{equation}

    \vspace{10pt}
    We are finally left with
    \begin{equation} \label{eq:vv_stress_tensor}
        \ev{T_{vv}}\big|_{r=0} = \ev{T_{vv}^{\text{(min)}}}\big|_{r=0} + \dfrac{\uv^2}{L^2} \,  T^{\text{(non-min)}}_{vv}[\Phi = \Phi^{\text{sol}}_{\text{min}}] \big|_{r=0} + \mathcal{O}(\hbar^2) \, .
    \end{equation}
    \vspace{2pt}

    \subsubsection{Generalized second law: a proof} \label{sec:proof}
    Using \eqref{eq:vv_stress_tensor}, the $vv$ component of the semi-classical equation of motion in \eqref{eq:EoM_metric} becomes
    \begin{equation} \label{eq:vv_EoM}
        E_{vv} \big|_{r=0} = 8\pi G_N \left[ \ev{T_{vv}^{\text{(min)}}} + \dfrac{\uv^2}{L^2} T_{vv}^{\text{(non-min)}}(\Phi_{\text{min}}^{\text{sol}}, D_a \Phi_{\text{min}}^{\text{sol}}, \ldots) \right]_{r=0} \, .
    \end{equation}
    Here $E_{vv}$ gets contribution only from $\act_{\text{g}}$ of the action \eqref{eq:action} while $T_{vv}^{\text{(non-min)}}$ is the classical equation of motion obtained from $\act_{\text{non-min}}$. We now invoke the $E_{vv}$ structure at $r=0$ in \eqref{eq:Evvfinal} \cite{Bhattacharyya:2021jhr,Biswas:2022grc}. Thus, \eqref{eq:vv_EoM} becomes
    \begin{equation}\label{eq:gslfinalstat}
        \partial_v \Theta_{\text{tot}} = 8\pi G_N \, \ev{T_{vv}^{\text{(min)}}} \, ,
    \end{equation}
    where $\Theta_{\text{tot}}$ is given by
    \begin{equation}\label{eq:Thetatotal}
        \Theta_{\text{tot}} =  \dfrac{1}{\sqrt{h}} \partial_v \left( \sqrt{h} \, \left[ \mathcal{J}^v + \dfrac{\uv^2}{L^2} \mathcal{J}^v_{\text{non-min}} \right] \right) + \dfrac{1}{\sqrt{h}} \partial_A \left( \sqrt{h} \, \left[ \mathcal{J}^A + \dfrac{\uv^2}{L^2} \mathcal{J}^A_{\text{non-min}} \right] \right) \, .
    \end{equation}
    Here $\mathcal{J}^v_{\text{non-min}}$ and $\mathcal{J}^A_{\text{non-min}}$ are given by \cite{Biswas:2022grc}
    \begin{equation}
        -T^{(\text{non-min})}_{vv} \big|_{r=0} = \partial_v \left( \dfrac{1}{\sqrt{h}} \partial_v \left( \sqrt{h} \, \mathcal{J}^v_{\text{non-min}} \right) + \dfrac{1}{\sqrt{h}} \partial_A \left( \sqrt{h} \,\mathcal{J}^A_{\text{non-min}} \right) \right) + \mathcal{O}(\hbar^2) \, ,
    \end{equation}
    and $\mathcal{J}^v$ and $\mathcal{J}^A$ are defined through \cite{Bhattacharyya:2021jhr}
    \begin{equation}\label{eq:evvstructure}
        E_{vv} \big|_{r=0} = \partial_v \left( \dfrac{1}{\sqrt{h}} \partial_v \left( \sqrt{h} \, \mathcal{J}^v \right) + \dfrac{1}{\sqrt{h}} \partial_A \left( \sqrt{h} \,\mathcal{J}^A \right) \right) + \mathcal{O}(\hbar^2)\, .
    \end{equation}
    \eqref{eq:gslfinalstat} implies a linearized GSL following the analysis of \cite{Wall:2011hj,Sarkar:2013swa}. The way one argues is as follows. In \eqref{eq:gslfinalstat}, we have split the minimal and non-minimal parts of the stress tensor. The part of the stress tensor that sits inside the expectation value is the minimal term, and the path integral is weighted with the minimal action. The matter fields of the full non-minimal theory might not always satisfy the Stability axiom of \cite{Wall:2011hj}. It is reasonable to assume that the algebra satisfies Determinism, Ultra-locality, and Local Lorentz symmetry. We assume that given such an algebra, we can consistently restrict the algebra to the minimal sector of the theory. This algebra is taken to satisfy the stability axiom
    as shown in \cite{Wall:2011hj}. Once we have this restricted ``minimal'' algebra of observables on the horizon, we can borrow the setup of section 5 of \cite{Sarkar:2013swa} to argue for the GSL.

    Using the axioms of local Lorentz symmetry and stability, we invoke the Bisongano-Wichmann theorem \cite{Bisognano:1976za}, which states that the vacuum state restricted to the Rindler wedge is thermal with respect to the boost energy. Then, we use Sewell's generalization \cite{Sewell:1982zz} to the algebra of observables on the horizon when the vacuum state is restricted to the horizon cut above the bifurcation surface. The vacuum then remains thermal for any cut $v > v'$ by virtue of the local Lorentz symmetry (supertranslations in this case). One can then use the monotonicity property of the relative entropy to argue for a GSL \cite{Wall:2011hj}. We note that this analysis assumes a suitable renormalization scheme, which maybe argued for based on Wall's original formalism \cite{Wall:2011hj} and the recent works in \cite{Chandrasekaran:2022cip, Chandrasekaran:2022eqq, Ali:2023bmm}.

\section{Discussion} \label{sec:discussion}
     Here, we summarize what we have been able to achieve so far. We considered arbitrary non-minimally coupled diffeomorphism invariant matter-gravity theories and established a linearized GSL for dynamical black holes in such theories. The non-trivial physics of non-minimally coupled theories is that it violates the Null Energy Condition. It is also unknown if they satisfy the Averaged Null Energy Condition (ANEC) \cite{Kontou:2020bta}. Thus, one cannot impose the stability axiom of \cite{Wall:2011hj} to help us in the proof. We got around this thorny issue by treating the higher derivative non-minimally coupled terms in an EFT sense. This allowed us to directly evaluate the matter path integral for the expectation value of the stress tensor. In this perturbative regime, it is clear that the non-minimal couplings only contribute to the horizon entropy via \eqref{eq:gslfinalstat}. With our formalism, it is not possible to treat the $S_{\text{QFT}}$ corrections to the generalized entropy from non-minimal terms since they are parametrically suppressed. \eqref{eq:vv_stress_tensor} implies that $S_{\text{out}}$ receives contributions only from the minimal sector. Our proof relied on some basic assumptions about the algebra of observables as in \cite{Wall:2011hj, Sarkar:2013swa}. It also suggests that the validity of the linearized GSL necessitates the perturbative EFT treatment of the higher derivative terms. We can then incorporate higher-order contributions. This is also in line with the recent observations of \cite{Hollands:2022fkn}, where the second law at the quadratic order in perturbations was valid only in the EFT sense.
    
    At this point, it is helpful to comment on how our proof is related to or different from the proofs of \cite{Wall:2010cj, Wall:2011hj, Sarkar:2013swa, Wall:2015raa, Ali:2023bmm}. \cite{Wall:2011hj} building on \cite{Wall:2010cj} worked in a regime where the matter fields across the horizon are rapidly fluctuating. However, the proof of \cite{Wall:2011hj} was restricted to Einstein's gravity, with matter fields minimally coupled to gravity. Our proof works in the regime of linearized fluctuations around stationary black holes, which is the same regime considered in \cite{Sarkar:2013swa, Wall:2015raa}. The proof outlined in \cite{Wall:2015raa} building on \cite{Sarkar:2013swa} was established for arbitrary pure gravity diffeomorphism invariant theories but with a minimally coupled matter sector only. We have been able to extend the proof of the GSL in \cite{Wall:2015raa} for arbitrary diffeomorphism invariant theories of gravity with matter fields non-minimally coupled to gravity, provided we are working in an EFT using the proof of the linearized second law worked out in \cite{Bhattacharya:2019qal, Bhattacharyya:2021jhr, Biswas:2022grc}. Further, unlike the case of \cite{Ali:2023bmm}, we have made a stronger statement of GSL in that we can show that the generalized entropy is increasing for every horizon cut, using the entropy current structure detailed in \cite{Bhattacharya:2019qal, Bhattacharyya:2021jhr, Biswas:2022grc}. Our proof crucially uses the advantages of working in the horizon adapted coordinates \eqref{eq:gauge} and the resulting entropy current structure for the $E_{vv}$ in \eqref{eq:evvstructure}. This significantly simplifies the analysis of the $S_{\text{out}}$, the entanglement entropy of the matter fields outside the horizon. The entropy current structure \eqref{eq:evvstructure} indicates that in \eqref{eq:vv_EoM}, the non-minimally coupled matter fields at one loop continue to contribute to the entropy of the horizon $S_{H}$ as expected classically. This was possible because of the EFT regime we worked with in the above analysis. This completes our proof of the linearized GSL for a matter sector that can be non-minimally coupled.

     We now highlight some future directions. Our analysis does not consider the graviton contribution. \cite{Ali:2023bmm} have taken the gravitons into account and it would be interesting to see how these subtleties fit into our perturbative EFT analysis. We also assume a sharp suppression of scale between minimal and non-minimal sectors, making our analysis not the most generic treatment of the non-minimal sector. Two crucial open issues remain: first, what happens when the non-minimal contribution is comparable to the minimal sector? In particular, we do not know how to treat all possible non-minimal couplings that could appear at two derivatives in the EFT treatment; second, our analysis only holds in the perturbative setup, and we do not yet understand how to incorporate any non-perturbative effects up to the linear order. 
     
     To conclude the discussion, we mention that the horizon entropy in our context is coarse-grained as we do not know its microstate description. It would be interesting to see if there is a holographic entanglement entropy analog of the monotonic quantity, as proposed in Section \ref{sec:proof},\footnote{This is in anticipation of an analogous calculation of the holographic entanglement entropy of Dong \cite{Dong:2013qoa} which agrees with the quantity satisfying the linearized semi-classical GSL in \cite{Wall:2015raa}.} that satisfies the linearized semi-classical GSL for higher curvature gravity theory with non-minimally coupled matter sector, and check whether the horizon will turn out to be the \textit{quantum extremal surface} (QES), \cite{Engelhardt:2014gca}, in that context. We do not expect the latter to hold; otherwise, it would correspond to saying that the horizon entropy is fine-grained in our setup. However, this is not the case as here entropy is derived from Wald-like prescription \cite{Iyer:1994ys} which in turn arises from $\delta M = T \delta S$, where $\delta M$ is given by the energy integral on the surface at infinity and thus represents the collective energy of the spacetime. It is an extensive quantity, and in this sense, the horizon entropy is coarse-grained. This could have some interesting connections to other recent works that rely on the QES formalism \cite{Engelhardt:2017aux, Bousso:2019dxk}.

\ack 
    We would like to thank Sayantani Bhattacharyya, Nilay Kundu, Alok Laddha, Sudipta Sarkar, Amitabh Virmani, Aron Wall, and Zihan Yan for helpful discussions at various stages of the project. Also, we thank A P Balachandran, Jyotirmoy Bhattacharya, Nilay Kundu, Sudipta Sarkar, Suneeta Vardarajan and Amitabh Virmani for several useful comments on the draft. This research was supported in part by the International Centre for Theoretical Sciences (ICTS) for the program ``Nonperturbative and Numerical Approaches to Quantum Gravity, String Theory and Holography" (code: ICTS/numstrings-2022/8). PD duly acknowledges the Council of Scientific and Industrial Research (CSIR), New Delhi, for financial assistance through the Senior Research Fellowship (SRF) scheme. KJ would like to thank the Department of Atomic Energy, Govt. of India, for financial support.

\appendix

\section{Examples of solutions in the horizon adapted coordinates}    \label{ap:examplegauge}
   This appendix will show how one can obtain the standard Schwarzschild and Kerr solutions in the horizon adapted coordinates of \eqref{eq:horizon_adapated_gauge}. Since we are using geodesics to rule our coordinates, the coordinate patch is valid only until the geodesics form a caustic. This necessarily means that the coordinate patch is valid only locally near the dynamical horizon. Our setup is akin to the \textit{Gaussian Null Coordinates} discussed in Section 2.1 of \cite{Kunduri:2013gce}. To expound on this point, we briefly demonstrate the following: we will show that the Schwarzschild metric can be bought to the gauge given by \eqref{eq:horizon_adapated_gauge} such that the metric is valid for arbitrary values of $r$ away from the horizon. However, in case of the Kerr metric we bring it to the gauge given by \eqref{eq:horizon_adapated_gauge} close to the horizon in a perturbative expansion only.

   \subsection*{Schwarzschild solution}
   Consider the Eddington-Finkelstein Patch of the Schwarzschild solution
    \begin{equation}
        ds^2 = 2 \dd v\, \dd r - \left(1 - \dfrac{2M}{r} \right) \dd v^2 + r^2 \dd \Omega^2 \, .
    \end{equation}
    $v,r$ are not affine and $r=0$ is not the horizon. Thus, do the following transformations successively:
    \begin{equation}
        v = 4M \, \log \lambda ~~ \rightarrow ~~ \tilde{r}=r-2M  ~~ \rightarrow ~~ \tilde{r}=\dfrac{\lambda \, \rho}{4M} \, .
    \end{equation}
    We can thus bring the metric to the desired gauge \eqref{eq:horizon_adapated_gauge} where $\lambda,\rho$ are affine and $\rho=0$ is the horizon:
    \begin{equation}
        ds^2 = 2 \dd \lambda\,\dd \rho + \rho^2 \left( \dfrac{2}{\lambda \rho} - \left(1 - \dfrac{8M^2}{\lambda\rho + 8M^2} \right)\dfrac{16 M^2}{(\lambda\rho)^2} \right) \dd \lambda^2 + \left(\dfrac{\lambda\rho}{4M}+2M\right)^2 \dd \Omega^2 \, .
    \end{equation}
    It is important to note that this metric is valid for arbitrary values of $\rho$ away from the horizon. Near the horizon $\rho=0$, this metric takes the desired form as,
    \begin{equation}
        ds^2 = 2 \dd\lambda\,\dd \rho + \rho^2 \left( \dfrac{1}{4} - \dfrac{\lambda\rho}{32M} + \dfrac{(\lambda\rho)^2}{256M^2} - \mathcal{O}[(\lambda\rho)^3] \right)\,\dd \lambda^2 + \left(\dfrac{\lambda\rho}{4M}+2M\right)^2 \dd \Omega^2 \, .
    \end{equation}

   \subsection*{Kerr solution}
   Consider the Eddington-Finkelstein patch of the Kerr solution
    \begin{equation}\label{Kerr}
        \begin{split}
            ds^2 = 2 \dd v\,\dd r - \left(1 - \dfrac{2Mr}{(r^2+a^2 \cos^2\theta)} \right) \dd v^2 -\dfrac{4Mar \sin^2\theta}{r^2+a^2 \cos^2\theta} \dd v\,\dd \phi + (r^2 + a^2 \cos^2\theta)\, \dd \theta^2 \\ - 2a \sin^2\theta \,\dd r\,\dd \phi
            + \dfrac{\sin^2\theta}{r^2 + a^2 \cos^2\theta} [ (r^2+a^2)^2 - (r^2 + a^2 - 2Mr)a^2 \sin^2 \theta ]\, \dd \phi^2 \, .
        \end{split}
    \end{equation}
    Following the arguments of Appendix A of \cite{Booth:2012xm}, we can bring the Kerr metric \eqref{Kerr} to our gauge. First, go to a co-rotating frame
    \begin{equation}
        \phi = \varphi + \dfrac{a}{r^2_{+}+a^2} v
    \end{equation}
    Re-write the metric with respect to the following null geodesics
    \begin{equation}
         \begin{split}
             n &= -\left(\dfrac{a^2 \sin^2 \theta }{2 r^2_{+} + 2 a^2 \cos^2 \theta}\right) \dfrac{\partial}{\partial v} - \left( \dfrac{r^2_{+}+a^2}{r^2_{+} + a^2 \cos^2 \theta } \right) \dfrac{\partial}{\partial r} - \left( \dfrac{a}{r^2_{+} + a^2} \right) \dfrac{\partial}{\partial \phi} \, , \\
             l &= \dfrac{\partial}{\partial v} \, .
         \end{split}
    \end{equation}
    Consider a family of null geodesics with $n$ as the tangent field crossing the horizon $H$ at the point $(v,\theta,\phi)$. Parametrize them using an affine $\rho$ such that $\rho = 0$ is $H$. Thus, we can perturbatively construct the geodesic coordinates: 
    \begin{equation}
        X^{\alpha}_{(v,\theta,\phi)}(\rho) \approx X^{\alpha}|_{\rho=0} + \rho \left. \dfrac{d \, X^{\alpha}}{d \, \rho} \right|_{\rho=0} + \dfrac{\rho^2}{2} \left. \dfrac{d X^{\alpha}}{d \rho} \right|_{\rho=0} + \mathcal{O}(\rho^3) \, ,
    \end{equation}
    \begin{equation}
        X^{\alpha}|_{H} = (v,r_{+},\theta,\phi) \, ,
    \end{equation}
    \begin{equation}
        \left. \dfrac{d X^{\alpha}}{d \rho} \right|_{H} = n^{\alpha} = \left(-\dfrac{a^2 \sin^2 \theta }{2 r^2_{+} + 2 a^2 \cos^2 \theta},-\dfrac{r^2_{+}+a^2}{r^2_{+} + a^2 \cos^2 \theta },0,-\dfrac{a}{r^2_{+} + a^2} \right) \, ,
    \end{equation}
    \begin{equation}
        n^{\beta} D_{\beta} n^{\alpha} = 0 \implies \left. \dfrac{d^2 X^{\alpha}}{d \rho^2} \right|_{H} = - \Gamma^{\alpha}_{\beta\gamma} n^{\beta} n^{\gamma} \, .
    \end{equation}
    Thus, we have the transformation from $(v,r,\theta,\phi)$ to $(v,\rho,\theta,\phi)$. The Kerr metric can thus be schematically written as
    \begin{equation}
        g_{\alpha\beta} \approx g^{(0)}_{\alpha\beta} + \rho \, g^{(1)}_{\alpha\beta} + \dfrac{\rho^2}{2} \, g^{(2)}_{\alpha\beta} + \mathcal{O}(\rho^3) \, .
    \end{equation}
    Once we reach (A.11) of \cite{Booth:2012xm}, we simply do the following transformations:
    \begin{equation}
        \rho = \kappa\, u\, \lambda, ~~~ v = \dfrac{1}{\kappa}\, \log \lambda, ~~~ \text{where } ~~~ \kappa = \dfrac{\Delta'}{2 \chi_o} \, ,
    \end{equation}
    where $u,\lambda$ are the new affine coordinates we define, and other quantities are defined as in \cite{Booth:2012xm}. This will result in the Kerr metric in our gauge \eqref{eq:horizon_adapated_gauge}:
    \begin{equation}
            \begin{split}
                ds^2 &= 2 \dd \lambda \, \dd u + \dfrac{u^2}{2} g^{(2)}_{vv} \, \dd \lambda^2 + u \, (2 g^{(1)}_{v\theta}+ \kappa u \lambda \, g^{(2)}_{v\theta}) \, \dd \lambda \, \dd \theta \\
                &+ u \, (2 g^{(1)}_{v\phi} + \kappa u \lambda) \, \dd \lambda \, \dd \phi + \left( g^{(0)}_{\theta\theta} + \kappa u \lambda g^{(1)}_{\theta\theta} + \dfrac{(\kappa u \lambda)^2}{2} g^{(2)}_{\theta\theta}\right) \, \dd \theta^2 \\
                &+\left( g^{(0)}_{\phi\phi} + \kappa u \lambda g^{(1)}_{\phi\phi} + \dfrac{(\kappa u \lambda)^2}{2} g^{(2)}_{\phi\phi} \right) \, \dd \phi^2 \\
                &+ \left( \kappa u \lambda g^{(1)}_{\theta\phi} + \dfrac{(\kappa u \lambda)^2}{2} g^{(2)}_{\theta\phi} \right) \, \dd \theta \, \dd \phi + \dots \, .
            \end{split}
        \end{equation}
    From the above analysis, we can see that the Kerr metric in our gauge \eqref{eq:horizon_adapated_gauge} is valid only close to the horizon.

\section*{References}
\bibliographystyle{myunsrt}
\bibliography{GSLref}

\end{document}